\documentstyle[12pt]{article}
\begin{document}
\textwidth=16cm
\textheight=22cm
\def\lb{\nextline}
\def\Order#1{{\cal O}($#1$)}
\def\alphas{$\alpha_s(M_{^Z})$}
\def\sovem{{s\over m^2_e}}
\def\Born{{\rm Born}}
\def\nubar{\bar{\nu}}
\def\ee{$e^+e^-$}
\def\nubart{\bar{\nu}_\tau}
\def\sstrut{$\strut\atop\strut$}

  \def\PLB #1 #2 #3 {{\it Phys. Lett.} {\bf {#1}B}	(#2)  #3}
  \def\NPB #1 #2 #3 {{\it Nucl. Phys.} {\bf B#1}	(#2)  #3}
  \def\PRD #1 #2 #3 {{\it Phys. Rev.} {\bf D#1}		(#2)  #3}
  \def\PP #1 #2 #3 {{\it Phys. Rep.} {\bf#1}		(#2)  #3}
  \def\PRL #1 #2 #3 {{\it Phys. Rev. Lett.} {\bf#1}	(#2)  #3}
  \def\CPC #1 #2 #3 {{\it Comp. Phys. Commun.} {\bf#1}	(#2)  #3}
  \def\ANN #1 #2 #3 {{\it Annals of Phys.} {\bf#1}	(#2)  #3}
  \def\APPB #1 #2 #3 {{\it Acta Phys. Polonica} {\bf B#1}(#2) #3}
  \def\ZPC #1 #2 #3 {{\it Zeit. f. Phys.} {\bf C#1}	(#2)  #3}
  \def\CPC #1 #2 #3 {{\it Comp. Phys. Commun.} {\bf#1}	(#2)  #3}
  \def\SJNP  #1 #2 #3 {{\it Sov. J. Nucl. Phys.} {\bf#1}(#3)  #3}
  \def\YadF  #1 #2 #3 {{\it Yad. Fiz.} {\bf#1}		(#2)  #3}

\def\uncatcodespecials{\def\do##1{\catcode`##1=12 }\dospecials}
\def\setupverbatim{\tt
  \def\par{\leavevmode\endgraf} \catcode`\`=\active
  \obeylines \uncatcodespecials \obeyspaces \parindent=5mm \parskip=0pt}
{\obeyspaces\global\let =\ } 
{\catcode`\`=\active \gdef`{\relax\lq}}
\def\beginverbatim{\par\begingroup\setupverbatim\doverbatim}
{\catcode`\|=0 \catcode`\\=12 
  |obeylines|gdef|doverbatim^^M#1\endverbatim{#1|endgroup}}
\begin{titlepage}

\begin{flushright}
\vbox{
{\bf TTP 92-37}\\
{\rm 23 Nov. 1992}\\
 }
\end{flushright}

\vspace{1.0cm}
\begin{center}
{\bf\Large LIGHT GLUINOS IN $Z^{_0}$ DECAYS ?}
\end{center}
\vspace{0.5cm}
\begin{center}
{\bf\Large Marek Je\.zabek}
   \footnote{Alexander von Humboldt Foundation Fellow.
   Permanent address:
    Institute of Nuclear Physics, Cracow, Poland,
    e-mail: JEZABEK@CHOPIN.IFJ.EDU.PL and BF08@DKAUNI2.BITNET}
{\bf\Large and  Johann H. K\"uhn}\\
    \vskip0.3cm
   {\it Institut f\"ur Theoretische Teilchenphysik}\\
   {\it Universit\"at Karlsruhe}\\
   {\it Kaiserstr. 12, Postfach 6980}\\
   {\it 7500 Karlsruhe 1, Germany}
\end{center}
\vskip1.0cm
\begin{center}
{\bf Abstract}
\end{center}
\begin{quote}
{\rm
We point out that an apparent discrepancy between the values of
\alphas\ as determined from low versus high energy experiments
can be explained if an electrically neutral coloured fermion
exists which slows down the running of the strong coupling constant
$\alpha_s$.}
\end{quote}

\end{titlepage}
      20 years after it was born QCD is commonly accepted as the
theory of strong interactions \cite{Aachen}. Two years of data taking
by LEP experiments contributed a lot to this belief. The impact of
\ee experiments on QCD can be compared only to the impact of deep
inelastic lepton-hadron  scattering (DIS). It is therefore worrying that
these two processes seem to imply slightly different values of
\alphas, c.f. Table 1 taken from {\cite{Bethke}}. The results
of deep inelastic lepton-nucleon scattering, crystal clear from
the theoretical point of view, are systematically below the results
of \ee analyses of event shapes. The discrepancy is even more striking
when {\alphas = 0.112 $\pm$ 0.004} as determined from DIS is compared
to the result of the improved analysis of event shapes, including
resumation of double logs and removing the unpleasant scale
dependence \cite{Catani,Webber}. The averaged LEP value obtained
from this analysis is {\alphas = 0.124 $\pm$ 0.005} \cite{Bethke}.
The data on \ee cross sections for annihilation also indicate
higher values of $\alpha_s$, albeit with much larger errors.

\begin{table}[h]
\begin{tabular}{||c|c|c|c||} \hline
Process       &  Q [GeV]   &   $\alpha_s(Q)$    & \alphas  \\   \hline
DIS [$\nu$]   &  $5.0$&  $0.193\pm\ {}^{.019}_{.018}$&$0.111\pm.006$\\
DIS [$\mu$]     &   $7.1$    & $0.180\pm{.014}$ & $0.113\pm.005$\\
\hline
\ee [ev.shapes] \\  \hline
PETRA,PEP,TRISTAN& $35.0$   & $0.140\pm.020$   & $0.119\pm.014$\\
AMY              & $58.0$   & $0.130\pm.008$   & $0.122\pm.007$\\
LEP,SLC          & $91.2$   & $0.120\pm.006$   & $0.120\pm.006$\\
\cline{2-4}
LEP[resumed]     & $91.2$   & $0.124\pm.005$   & $0.124\pm.005$\\
\hline
\ee [$\sigma$]  \\   \hline
\ee[$\sigma_{had}$]& $34.0$   & $0.157\pm.018$   & $0.131\pm.012$\\
$\Gamma(Z^0\rightarrow had)$& $91.2$   & $0.130\pm.012$   &
$0.130\pm.012$\\
\hline\hline
PDG'92           &        &                & $0.1134\pm.0035$ \\
\hline
\end{tabular}
\caption{Strong coupling constant \alphas\ determined from
deep inelastic lepton-nucleon and \ee annihilations. The
experimental values are taken from ref.[2]~.}
\end{table}

A possible and likely explanation of the above mentioned problem
is that the theoretical uncertainties and the related systematic
errors are under\-estimated in the analyses \cite{Altarelli}.
However, an alternative solution exists which should not be
overlooked until it is really excluded by experiment.\par
It is possible that an electrically neutral coloured fermion
of relatively low ($\sim$ a few GeV) mass slows down the running
of $\alpha_s$ between the scales accessible to the `high energy'
\ee and those being probed by `low energy' deep inelasting
experiments. An obvious candidate is a light gluino. \par
We are aware
of arguments, based on unification and cosmology, disfavouring
existence of light gluino; see also \cite{PDG92}~.
On the other hand we are impressed by the fact that, contrary to
widespread belief and earlier claims \cite{Ant}, the evolution
of $\alpha_s$ between 5 and 90 GeV is described much better by
the formula \cite{SUSY}
\begin{eqnarray}
&\alpha_s(Q) = {4\pi\over b_0 \ln Q^2/\Lambda^2}
\left[ 1 - {b_1\over {b_0}^2}
{\ln\ln Q^2/\Lambda^2\over \ln Q^2/\Lambda^2} \right] \\
& b_0 = 11 - {2\over 3}N_f - 2N_{\tilde g}, \qquad
 b_1 = 102 - {38\over 3}N_f - 48 N_{\tilde g} \nonumber
\end{eqnarray}
for $N_f$=5 quark flavours and $N_{\tilde g}$=1 gluino than for
$N_{\tilde g}$=0, see Table 2.

\begin{table}[h]
\begin{tabular}{||c|c|c|c||} \hline
Q [GeV]   &   $N_{\tilde g}=0$    & $N_{\tilde g}=1$ & Exp.  \\   \hline
$91.2$  & $0.124\pm.005$ & $0.124\pm.005$    & $0.124\pm.005$\\  \hline
$58.0$  & $0.134\pm.006$ & $0.130\pm.006$    & $0.130\pm.008$\\  \hline
$34.0$  & $0.147\pm.007$ & $0.139\pm.006$    & $0.140\pm.020$\\  \hline
$20.0$  & $0.164\pm.009$ & $0.148\pm.007$    &
$0.138 \pm\ {}^{.028}_{.019}$\\ \hline
$10.0$  & $0.193\pm.013$ & $0.164\pm.009$    &
$0.167\pm\ {}^{.015}_{.011}$\\    \hline
$ 7.1$  & $0.212\pm.015$ & $0.172\pm.010$    & $0.180\pm.014$\\  \hline
$ 5.0$  & $0.235\pm.019$ & $0.182\pm.011$    & $0.193\pm.019$\\
\cline{4-4}
        &                &                   & $0.174\pm.012$\\  \hline
\end{tabular}
\caption{Evolution of the strong coupling constant $\alpha_s(Q)$
for $N_{\tilde g}=$0 and 1 starting from the value at 91.2 GeV
down to 5 GeV.
The experimental values are taken from ref.[2]~.}
\end{table}

In conclusion we would like to stress that the existing data do not
preclude the possibility that a coloured and electrically neutral
fermion exists which slows down the running of $\alpha_s$ between
5 and 90 GeV. A search for, or exclusion of such a particle
in 4-jet \ee or 3-jet ep events should be persued by the experiments
at LEP and HERA.


\vskip 1cm
{\bf Acknowledgements}~. We thank K. Chetyrkin and H. M\"uller
for stimulating discussions. Work supported in part by Polish
Committe for Scientific Research and BMFT.

\end{document}